\documentstyle[12pt,epsfig]{article}
\begin{document}

\newcommand \be  {\begin{equation}}
\newcommand \bea {\begin{eqnarray} \nonumber }
\newcommand \ee  {\end{equation}}
\newcommand \eea {\end{eqnarray}}

\def\(({\left(}
\def\)){\right)}
\def\[[{\left[}
\def\]]{\right]}
\def\bi{\bibitem}
\def \la{\langle}
\def\ra{\rangle}
\def \a{\alpha}
\def\ov{\over}
\def\l{\left}
\def\r{\right}
\def\b{\beta}
\def\de{\delta}
\def\p{\partial}
\def\D{\Delta}
\def\eps{\epsilon}
\def\lam{\lambda}
\def\hl{{\hat \lambda}}

\def \v{ {\vec v}}
\def \vx{\vec x}

\title{\bf PHENOMENOLOGY OF THE INTEREST RATE CURVE.}

\vskip 3 true cm

\author{
 Jean-Philippe Bouchaud$^{1,2,\dagger}$, Nicolas Sagna$^{3,*}$, Rama Cont$^{2,1}$,\\
 Nicole El-Karoui$^4$ and Marc Potters$^2$}

\date{\it
$^1$ Service de Physique de l'\'Etat Condens\'e\\
 Centre d'Etudes de Saclay\\ 
91191 Gif-sur-Yvette Cedex, France \\ 
$^2$ Science \& Finance, 109-111 rue Victor-Hugo, 92532 Levallois, France \\
$^3$ Laboratoire de Probabilit\'es, URA CNRS 224\\ Universit\'e de Paris VI, 
Tour 56, 4 place Jussieu\\ 75252 Paris 05, France\\
$^4$ {\sc{cmpax}}, Ecole Polytechnique, 91 128 Palaiseau CEDEX. \\
$^\dagger$ Author to whom correspondence should be sent.\\
$^*$ Present address: Credit Suisse First Boston, Fixed Income Research Group\\
One Cabot Square
London E14 4QJ, U.K. }


\maketitle

\begin{abstract}

This paper contains a phenomenological description of the whole
U.S. forward rate curve ({\sc frc}), based on an data in the period 1990-1996. 
We find that the {\it average} {\sc frc} (measured from the spot rate) grows as the square-root of the maturity, with a prefactor which is comparable to the spot rate volatility. This suggests that forward rate market prices include a risk premium, comparable to the probable changes of the spot rate between now and maturity, which can be understood as a `Value-at-Risk' type of pricing. The {\it instantaneous} {\sc frc} however departs form a simple square-root law. The distortion is maximum around one year, and reflects the market anticipation of a local trend on the spot rate. This anticipated trend is shown to be calibrated on the past behaviour of the spot itself. We show that this is consistent with the volatility `hump' around one year found by several authors (and which we confirm). Finally, the number of independent components needed to interpret most of the {\sc frc}  fluctuations is found to be small. We rationalize this by showing that the dynamical evolution of 
the {\sc frc} contains a stabilizing second derivative (line tension) term, which tends to suppress short scale distortions
of the {\sc frc}. This shape dependent term could lead, in principle, to arbitrage. However,  this arbitrage cannot be implemented in practice because of transaction costs. We suggest that the presence of transaction costs (or other market `imperfections') is crucial for model building, for a much wider class of models becomes eligible to represent reality.

\end{abstract}

\vskip 0.5cm

\vskip 0.5cm

\noindent Electronic addresses : 
bouchaud@amoco.saclay.cea.fr\\
cont@ens.fr\\
elkaroui@cmapx.polytechnique.fr\\
nsagna@csfp.co.uk\\
potters@cfm.fr

\newpage
\section{Introduction}

Finding an adequate statistical description of the dynamics of financial assets has a
long history which is probably reaching a climax, now that enormous sets of time series
are readily available for analysis. The case of the interest rate curve is
particularly complex and interesting, since it is not the random motion of a point, but
rather the history of a whole curve (corresponding to different maturities) which is at
stake. The need of a consistent description of the whole interest rate curve is
furthermore enhanced by the rapid development of interest rate derivatives (options, swaps,
options on swaps, etc) \cite{hull}. Present models of the interest rate curve fall into two
categories \cite{Risk}: the first one is the Vasicek model and its variants, which focuses on the
dynamics of the short term interest rate, from which the whole curve is reconstructed (see \cite{rodgers} for a recent review).
The second one, initiated by Heath, Jarrow and Morton \cite{hjm} (see also \cite{HoLee}), takes the full forward
rate curve as 
dynamical variables, which is driven by one (or several) continuous Brownian motion,
multiplied by a maturity dependent scale factor.  Most models are 
primarily motivated by their mathematical tractability rather than by their
ability to describe the data. For example, the fluctuations are often assumed to be gaussian, thereby neglecting `fat tail' effects. More importantly, as we shall discuss below, the empirical {\it shape} of the interest rate curve can only be captured by these models if one includes an additional parameter, the
so called `market price of risk', which takes a rather large value.

The aim of this paper is threefold. We first present an empirical study of the 
forward rate curve ({\sc frc}), where we isolate several important features 
which a good model should be asked to reproduce. Rather than aiming at a precise
statistical determination of parameters (which is known to be particularly difficult
for all financial data), we wish to offer a qualitative description of the dynamics of the {\sc frc} and try to decipher the information contained in its shape, in terms of an `anticipated risk' and an `anticipated bias'. Based on
empirical evidence, we argue that the market fixes the interest rate curve through
a Value-at-Risk type of condition, rather than through an averaging procedure, which
is the starting point of the classical models alluded to above. Furthermore the `anticipated bias' is found to be strongly correlated with the past trend on the spot rate itself. We then present a
general class of string models, inspired from statistical physics, which describe
the motion of an elastic {\it curve} driven by noise, and discuss how these models
offer a natural framework to account for some of the empirical results, in particular
the  small number of independent factors needed to describe the evolution of the {\sc
frc}. Finally, we discuss the general concept of arbitrage and its relation with model
building. We suggest that the argument of absence of arbitrage opportunities, usually
used to restrict possible models of reality, is much too strong: in the presence of
transaction costs and/or residual risks, qualitatively new classes of models can be
considered.

As an important preliminary remark, we want to stress that the adequacy of a model 
can be assessed from two rather different standpoints. The
first one, which we shall adopt here, is to see how well a given
model describes the dynamics of the primary object, namely the forward rate curve.
The second one, more concerned with derivative markets, asks how consistent are the
determination of the model's parameters across several different derivative products,
irrespective of the ability of these parameters to reproduce the statistics of the
underlying asset. A well known example is the case of the Black-Scholes theory of
options: while there should be no volatility smile in a Black-Scholes world, it is
still an intersting question from an applied point of view to know whether the
implied volatility on plain vanilla options can be used to price (within a
Black-Scholes scheme) exotic options. We will however
not consider in the present paper the consequence of our model to the problem of
derivative pricing.

\section{Statistical analysis of the forward rate curve}

\subsection{Presentation of the data and notations}

The forward interest rate curve ({\sc frc}) at time $t$ is fully specified by the
collection of all forward rates $f(t,\theta)$, for different maturities $\theta$. It
allows for example to calculate the price $B(t,\theta)$ at time $t$ of a (zero-coupon)
bond, which pays $1$ at time $t+\theta$. It is given by: 
\be
B(t,t+\theta)= \exp - \int_0^{\theta} du \ f(t,u) \label{bonddef}
\ee
$r(t)=f(t,\theta=0)$ is
called the `spot rate'. Note that in the following $\theta$ is always a
time difference; the maturity {\it date} $T$ is $t+\theta$. 

\subsection{The data set}

Our study is based on a data set of daily prices of Eurodollar futures
contracts. 
The Eurodollar futures contract is a futures contract on an interest rate,
as opposed to the Treasury bill futures contract which is a futures contract
on the price of a Treasury bill. 
The interest rate underlying the Eurodollar futures contract is a 90-day rate,  earned on dollars deposited in a bank outside the U.S. by another bank. When interest rates are fixed, a well known arbitrage argument \cite{cir} implies that forward and futures contracts must have the same value.
But when interest rates are stochastic in principle  forward contracts and future contracts are no longer identical -- 
they have different margin requirements -- and one may expect slight differences
 between futures and forward contracts \cite{rendleman}, which we shall 
neglect in the following. We have furthermore checked that the effects discussed below are qualitatively the same as those observed when {\sc frc} is reconstructed from swap rates. The interest of studying forward rates rather 
than yield curves is that one has a direct access to a `derivative' (in the mathematical sense -- see Eq.(\ref{bonddef})), which obviously contains more
precise informations.

In practice, the futures markets price three months forward rates for 
{\it fixed} expiration dates, separated by three month intervals. Identifying three months futures rates
to instantaneous forward rates, we have available time series on forward rates
$f(t,T_i-t)$, where $T_i$ are fixed dates (March, June, September and December of each
year), which we have converted into fixed maturity (multiple of three months) forward
rates by a simple linear interpolation between the two nearest points such that 
$T_i - t \leq \theta \leq T_{i+1} - t$. Days where one contract disappears are not included, to remove possible artefacts. Between 1990 and 1996, we have at least 15 different
Eurodollar maturities for each market date; longer maturities are in general less liquid than short maturities; we however believe that such a difference in liquidity does not affect the qualitative conclusions drawn in this paper (this is partly confirmed by our analysis of the swap rates). Between 1994 and 1996, the number of available maturities rises to
30 (as time grows, longer and longer maturity forward rates are being traded on future
markets); we shall thus often use this restricted data set. Since we only have daily data, our reference time scale will be $\tau=1$
day. The variation of $f(t,\theta)$ between $t$ and $t+\tau$ will be denoted as
$\delta f(t,\theta)$: 
\be
\delta f(t,\theta) = f(t+\tau,\theta)-f(t,\theta)
\ee
It would obviously be very interesting to extend the present analysis to intra-day fluctuations. 

\subsection{Quantities of interest and data analysis}

The description of the {\sc frc} has two, possibly interrelated, aspects : 

(i) what is, at a given instant of time, the {\it shape} of the {\sc frc} as a
function of the maturity $\theta$ ?

(ii) what are the statistical properties of the increments $\delta f(t,\theta)$ between time $t$ and
time $t+\tau$, and how are they correlated with the shape of the {\sc frc} at time
$t$ ?

The two basic quantities describing the {\sc frc} at time $t$ is the value
of the short term interest rate  $f(t,\theta_{{\min}})$ (where $\theta_{\min}$ is the shortest available maturity), and that of the short term/long term {\it spread}
$s(t)=f(t,\theta_{\max})-f(t,\theta_{\min})$, where $\theta_{\max}$ is the longest
available maturity. The two quantities \footnote{We shall from now on take the three
month rate as an approximation to the spot rate $r(t)$.} $r(t) \simeq f(t,\theta_{{\min}}),\ s(t)$ are
plotted versus time in Fig. 1; note that 

-- The volatility $\sigma$ of the spot rate $r(t)$ is equal to 
\footnote{The
dimension of $r$ should really be $\%$ per year, but we conform to the habit of quoting $r$ simply in
$\%$. Note that this can sometimes be confusing when checking the correct dimensions 
of a formula.}
$0.8\%$/$\sqrt{\rm{year}}$. This is obtained by averaging over the whole period. However, as
previously noticed \cite{chan,Wilmott}, there seems to be a systematic correlation between $\sigma$ and $r$, which can be parametrized as $\sigma \propto r^\beta$. For the
period considered (1990-96), we find a much smaller value $\beta \simeq 0.4$ than the one reported in \cite{chan} $\beta =1.5$, based one monthly data from 1964 to 1989, or $\beta=1.1$ \cite{Wilmott}, based on weekly data from 1984 to 1993.

-- $s(t)$ has varied between 0.53\% and
4.34\%. Contrarily to some european interest rates on the same period, $s(t)$
has always remained positive, a situation we shall refer to as `normal'. (This however
does not mean that the {\sc frc} is increasing monotonously, see below).

\begin{figure}
\centerline{\hbox{\epsfig{figure=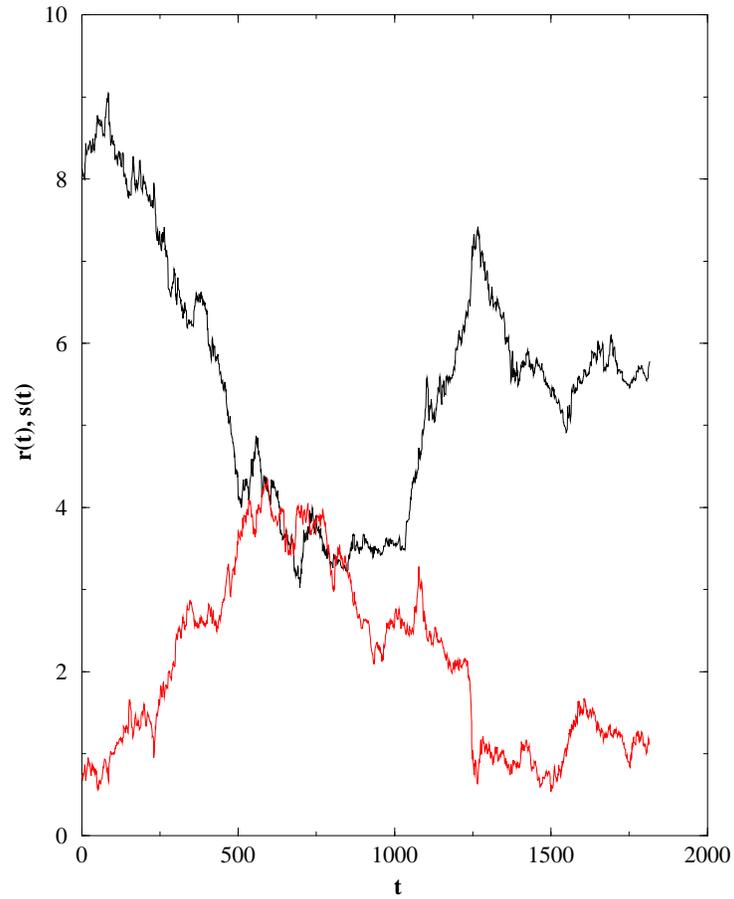,width=8cm}}}
\vskip 0.8cm
\caption{The historical time series of the spot rate $r(t)$ from 1990 to 1996 (top curve) -- actually
corresponding to  a three month future rate (dark line) and of the `spread' 
$s(t)$ (bottom curve), defined with the longest maturity available over the whole
period 1990-1996 on future markets, i.e. $\theta_{\rm max}=$ four years. (In restricted period (94-96), this
maturity grows to $\theta_{\rm max}=$  8 years.)} \label{fig1} \end{figure}

In order to rationalize the empirical data, we shall postulate that the whole 
{\sc frc} oscillates around an average shape, and {\it parametrize} $f(t,\theta)$ as
\footnote{The relation between this decomposition and the more standard Principal Component 
Analysis is clarified below. Note in particular that $r(t),\ s(t)$ and $\xi(t,\theta)$
are {\it a priori} not independent.}: 
\be
f(t,\theta) = r(t) + s(t) \frac{Y[\theta]}
{Y[\theta_{\max}]}+ \xi(t,\theta)\label{decomp1}
\ee
where $Y$ is a certain (time independent) function and $\xi(t,\theta)$ 
represents the deviation from the (empirical) average curve, in the sense that 
\be
\langle \xi(t,\theta) \rangle = 0
\ee
where $\langle ... \rangle$ represents a time average. By definition,
$\xi(t,\theta_{\min})=\xi(t,\theta_{\max}) \equiv 0$. 

Fig. 2 shows the scaling function $Y(u)$, determined by averaging 
the difference ${f(t,\theta)-r(t)}$ over time. Interestingly, the function $Y(u)$ is 
rather well fitted by a simple 
$\sqrt{u}$ law. This means that on average, the difference between the forward rate with maturity $\theta$ and the spot rate is equal to $A \sqrt{\theta}$, with a
proportionality constant $A=0.85 \%$/$\sqrt{\rm{year}}$ which turns
out to be nearly identical to the spot rate volatility.  We shall propose a simple
interpretation of this fact below.

\begin{figure}
\centerline{\hbox{\epsfig{figure=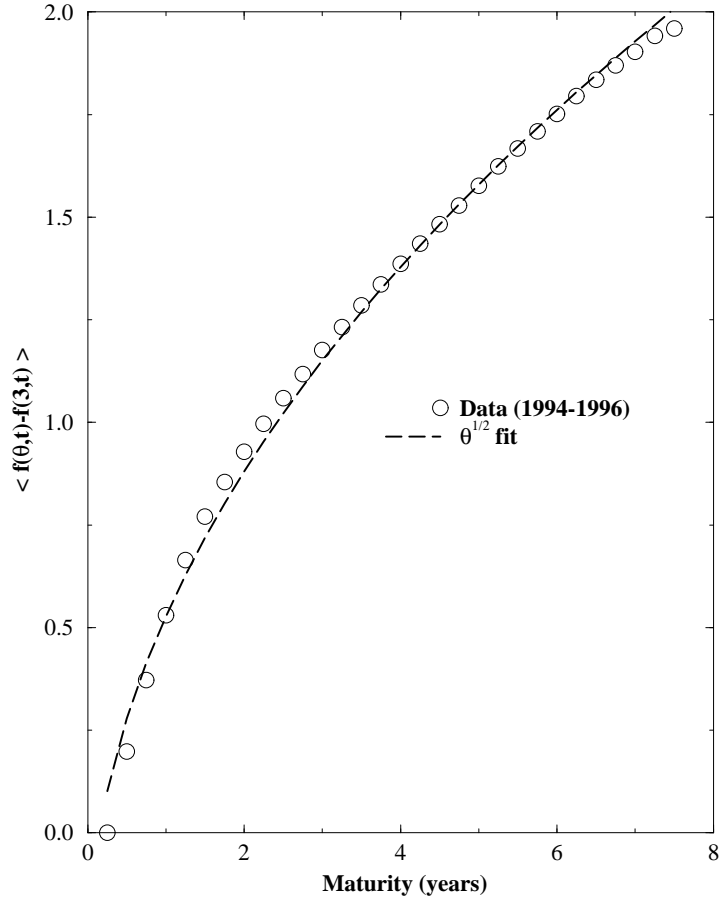,width=8cm}}}
\vskip 0.8cm
\caption{The average {\sc frc} in the period 94-96, as a function of the maturity 
$\theta$.
We have shown for comparison a one parameter fit with a square-root law,
$A(\protect\sqrt{\theta}- \protect\sqrt{\theta_{\min}})$. 
The same $\protect\sqrt{\theta}$ behaviour actually extends up to $\theta_{\max}=10$ years,
which is available in the second half of the time period.} \label{fig2}
\end{figure} 

Let us now turn to an analysis of the {\it fluctuations} around the average shape.
Fig. 3 shows for different instants of time the whole `deviation' $\xi(t,\theta)$ as a
function of $\theta$. 

\begin{figure}
\centerline{\hbox{\epsfig{figure=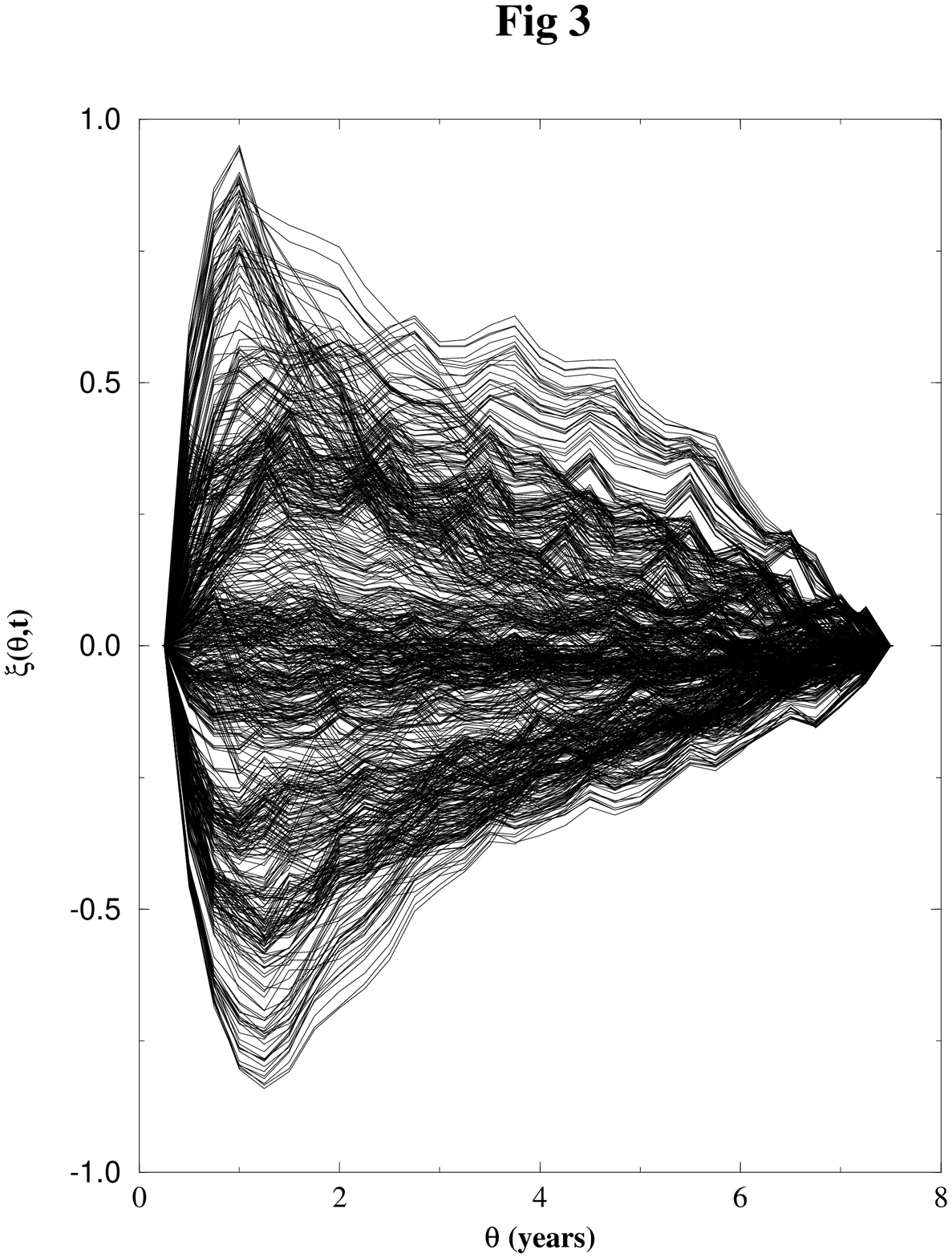,width=8cm}}}
\vskip 0.8cm
\caption{This shows the deviation $\xi(t,\theta)$ as a function of 
$\theta$ for different times $t$. Note that this resembles the motion of an elastic
string, with a maximum around $\theta$ = 1 year.} \label{fig3}
\end{figure}

These fluctuations are actually similar to that
of a vibrated elastic string. The average deviation $\Delta(\theta)$ can be defined 
as:
\be 
\Delta(\theta) := \sqrt{\langle \xi(t,\theta)^2 \rangle}
\ee
which is plotted in Fig. 4, for the period 94-96. The maximum of $\Delta$ is reached for a maturity of
$\theta^*=$1 year.

\begin{figure}
\centerline{\hbox{\epsfig{figure=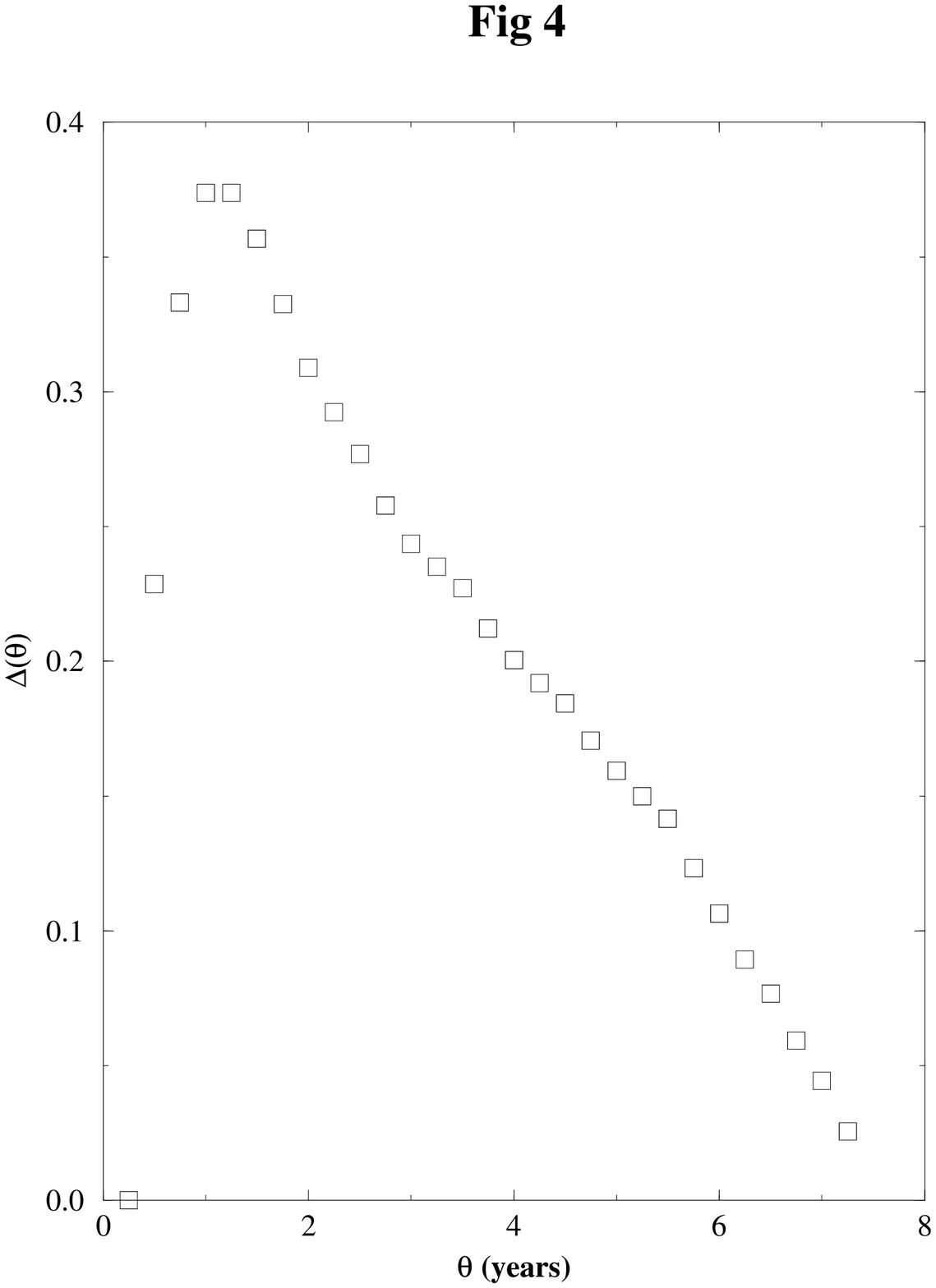,width=8cm}}}
\vskip 0.8cm
\caption{Root mean square deviation $\Delta(\theta)$ from the average 
{\sc frc} as a function of $\theta$. Note the maximum for $\theta^*=1$ year, for
which $\Delta \simeq 0.38 \%$.} 
\label{fig4}
\end{figure} 

A more precise study of the fluctuation modes
consists in studying the equal time correlation matrix ${\cal M}(\theta,\theta')$
defined as: 
\be
{\cal M}(\theta,\theta') = \langle \xi(t,\theta) \xi(t,\theta') \rangle
\ee
Note that ${\cal M}(\theta,\theta) \equiv \Delta(\theta)^2$. The eigenvalues $M_q^2$ of 
this matrix are plotted in Fig. 5, as a function of their
rank $q=1,2,...$, in log-log coordinates. We have shown for comparison a
$q^{-2}$ and $q^{-4}$ decay (see section 5), indicating that these eigenvalues are
decreasing very fast with $q$.  We have furthermore calculated the same eigenvalues in
the presence of some `artificial' noise (i.e. a random variable of zero mean and width
$0.04 \%$ added to each $f(t,\theta)$ independently). One sees that it only affects the
high $q$ modes, while leaving the modes with $q < 9$ relatively stable, and which
are thus statistically meaningful. The deviation from the average shape can  be
written as:  \be \xi(t,\theta) = \sum_q {M_q} \xi_q(t) \Psi_q(\theta) \ee where
$\langle \xi_q(t) \xi_{q'}(t) \rangle = \delta_{q,q'}$. Exactly as an elastic string,
the first mode $\Psi_1$ has no nodes (and is found to be nearly proportionnal to
$\Delta(\theta)$), while the second $\Psi_2$ has one node.

\begin{figure}
\centerline{\hbox{\epsfig{figure=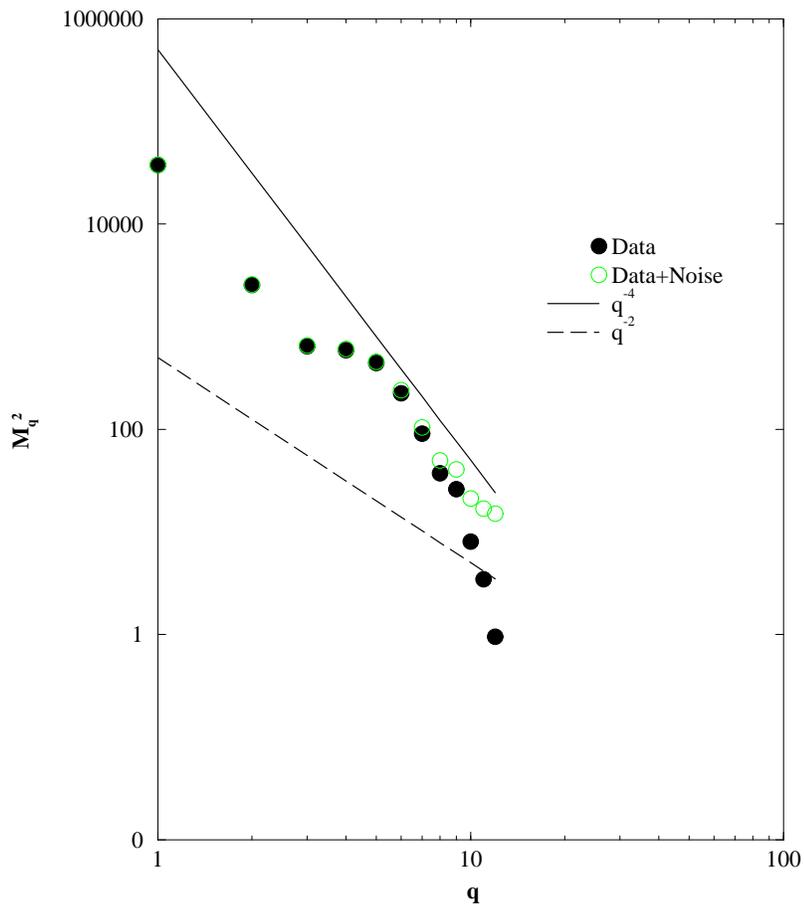,width=8cm}}}
\vskip 0.8cm
\caption{
The eigenvalues $M_q^2$ of the instantaneous fluctuation matrix 
${\cal M}(\theta,\theta')$, as a function of their
rank $q=1,2,...$, in log-log coordinates. We have shown for comparison a
$q^{-2}$ and a $q^{-4}$ decay (see section 5), indicating that these eigenvalues are decreasing very
fast with $q$. Also shown for comparison the eigenvalues of ${\cal M}(\theta,\theta')$
when an artificial noise (uniform in $[-0.02,+0.02]$) is independently added to each point of the {\sc frc}, to estimate the
relevance of the information contained in the high $q$ eigenvalues.
}
\label{fig5}
\end{figure} 

\subsection{Statistics of the daily increments}

We now turn to the statistics of the {\it daily} increments $\delta f(t,\theta)$ of 
the forward rates, 
by calculating their volatility $\sigma(\theta)=\sqrt{\langle \delta f(t,\theta)^2
\rangle}$, their excess kurtosis 
\be
\kappa(\theta) = \frac{\langle \delta f(t,\theta)^4 \rangle}{\sigma^4(\theta)} -3
\ee
and the following `spread' correlation function:
\be
{\cal C}(\theta) =\frac{\langle \delta f(t,\theta_{\min}) \left(\delta f(t,\theta)-\delta f(t,\theta_{\min})\right)\rangle}{\sigma^2(\theta_{\min})} \label{corr}
\ee
which measures the influence of the short term interest fluctuations on the other modes of motion of the {\sc frc}. 

Fig. 6 shows $\sigma(\theta)$ and $\kappa(\theta)$. Somewhat surprisingly,
$\sigma(\theta)$, much like $\Delta(\theta)$ has a maximum around $\theta^*$ = 1 year, a
feature already noticed in, e.g., \cite{hullwhite2,moraleda2}. The order of magnitude
of $\sigma(\theta)$ is $0.05 \%$/$\sqrt{\rm{day}}$, or $0.8\%$/$\sqrt{\rm{year}}$. 
The kurtosis $\kappa(\theta)$ is rather high (of the order of $5$), and only weakly 
decreasing with $\theta$. 

\begin{figure}
\centerline{\hbox{\epsfig{figure=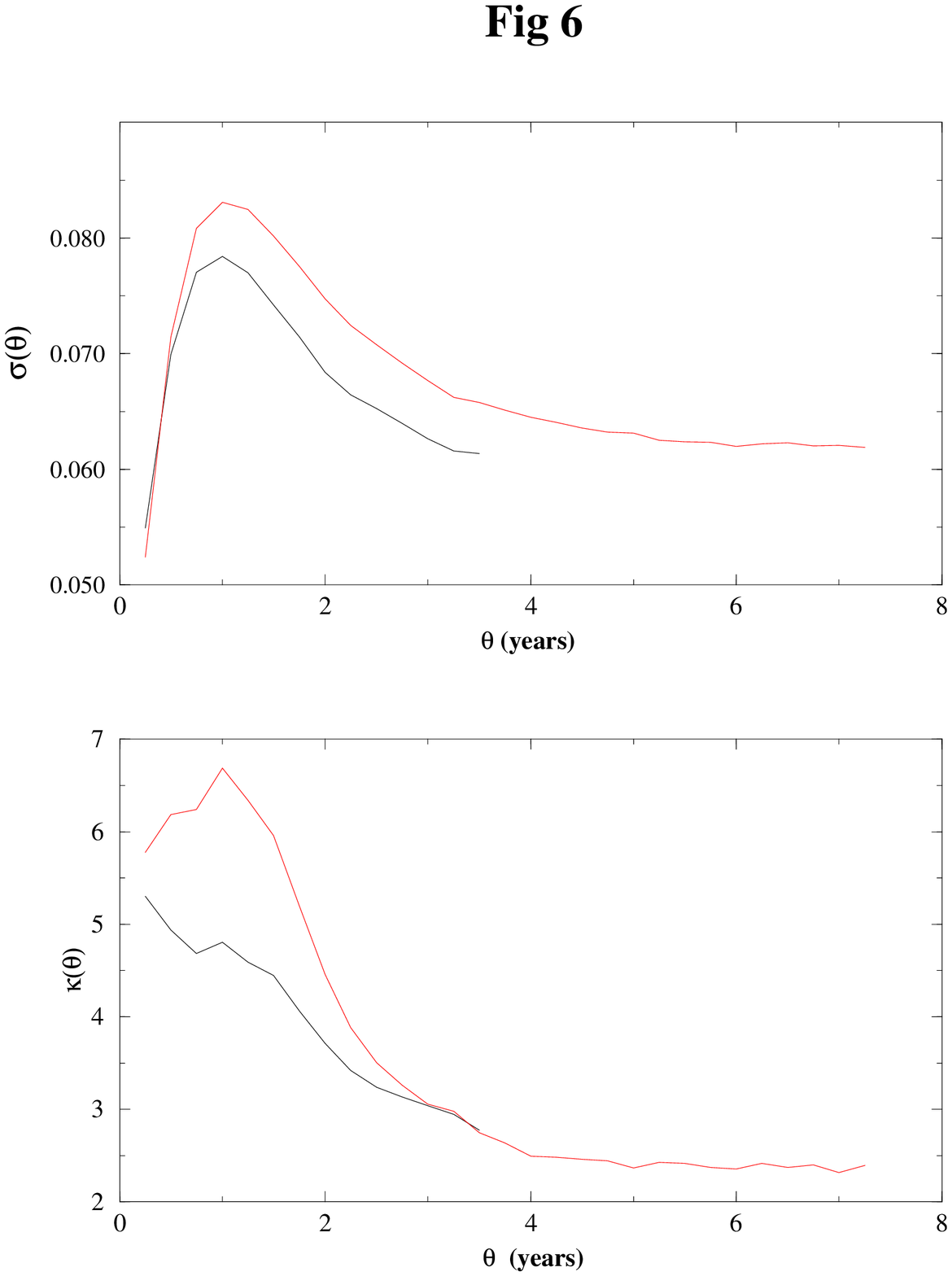,width=8cm}}}
\vskip 0.8cm
\caption{
The daily volatility and kurtosis as a function of maturity. Note the maximum of the 
volatility for $\theta=\theta^*$, while the kurtosis is rather high, and only very
slowly decreasing with $\theta$. The two curves correspond to the periods 90-96 and
94-96, the latter period extending to longer maturities. }
\label{fig6}
\end{figure}

Finally, ${\cal C}(\theta)$ is shown in Fig. 7; its shape is again very similar
to those of $\Delta(\theta)$ and $\sigma(\theta)$, with a pronounced
maximum around $\theta=1$ year. This means that the fluctuations of the 
short term rate are {\it amplified} for maturities around one year. We shall come back to this
important point below. Note that  ${\cal C}(\theta)$ goes to zero
for large maturities, which means that the short term rate and the short term/long term spread evolve, to a first approximation, in an uncorrelated
manner. A simple fit of ${\cal C}(\theta)$, shown in Fig. 7, is given by ${\cal C}(\theta)= c \theta \exp(-\Gamma \theta)$, with $c=1.5$ year$^{-1}$ and 
$\Gamma= 1.0$ year$^{-1}$. An interpretation of the values of $c$ and $\Gamma$
will be given below.

\begin{figure}
\centerline{\hbox{\epsfig{figure=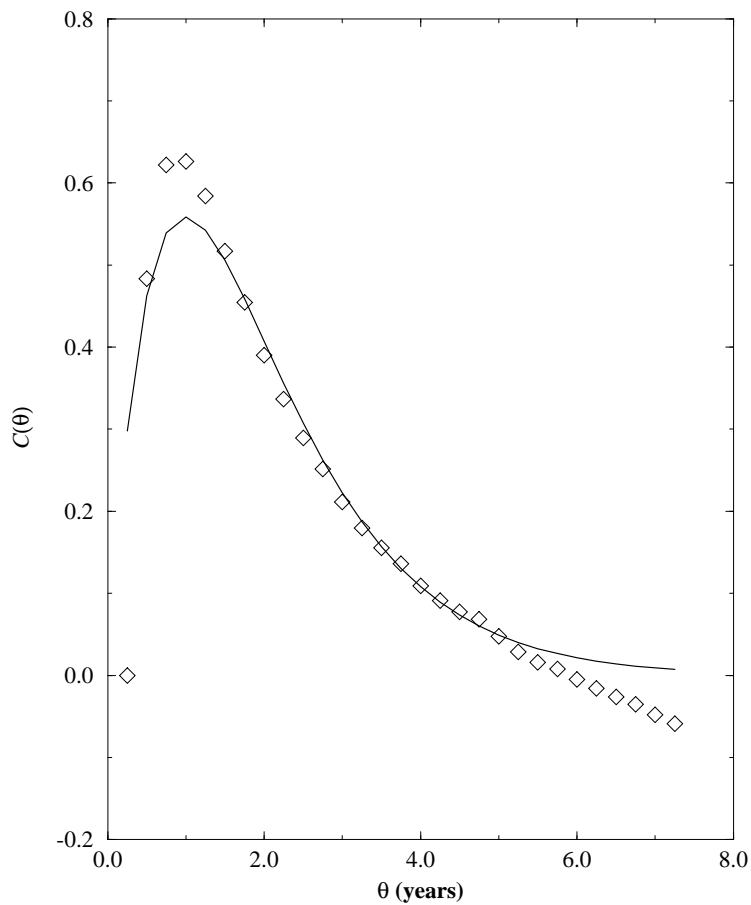,width=8cm}}}
\vskip 0.8cm
\caption{
The correlation function ${\cal C}(\theta)$ (defined by Eq.\  \protect\ref{corr}) between the daily variation of the
spot rate and that of the forward rate at maturity $\theta$, in the period 94-96. Again, ${\cal C}(\theta)$ is maximum for $\theta=\theta^*$, and decays rapidly beyond. A simple fit using the function $c \theta \exp(-\Gamma \theta)$
is shown for comparison, which leads to $c=1.5$ year$^{-1}$ and 
$\Gamma= 1.0$ year$^{-1}$. Note also that ${\cal C}(\theta_{\max})$ is close to
zero, indicating that, in a first approximation, the evolution of the spot rate
and of the spread are independent.}
\label{fig7}
\end{figure}

Much as above, one can define a correlation matrix 
for the variations of $\xi(t,\theta)$ as:
\be
{\cal N}(\theta,\theta') = \langle \delta \xi(t,\theta) \delta \xi(t,\theta') \rangle
\ee
Its diagonalisation gives results very similar to the
diagonalisation of ${\cal M}$, in particular concerning the fall-off of the eigenvalues as a function of $q$ and the shape of the low $q$
modes. 

Owing to the rapid decrease of the eigenvectors of ${\cal M}$ and ${\cal N}$, a good
approximation thus consists in retaining only the so called `butterfly' mode
of deformation $\Psi_1$, and to write \footnote{We have included the factor
${M_1}$ in the definition of $\Psi_1$}: 
\be
f(t,\theta) =  r(t) +  s(t) {\cal Y}\left[\frac{\theta}{\theta_{\max}}\right] + \xi_1(t) \Psi_1(\theta)
\label{mode1}
\ee
where ${\cal Y}(u):=\sqrt{u}$. That only a rather small degrees of freedom are needed to describe most of the {\sc frc}'s fluctuations was already discussed on several
occasions \cite{Longstaff,hullwhite2,schlogl,Douady}, although not exactly in the present terms.

\section{Classical models}

\subsection{Vasicek}

The simplest {\sc frc} model is a one factor model due to Vasicek \cite{vasicek}, where the whole term structure
can be ascribed to the short term interest rate, which is assumed to follow a
stochastic evolution described as:
\be
dr(t) = \Omega(r_0-r(t)) + \sigma dW(t)
\ee
where $r_0$ is an `equilibrium' reference rate, $\Omega$ describes the strength of the
reversion towards $r_0$ (and is the inverse of the mean reversion time), and $dW(t)$
is a Brownian noise, of volatility 1. In its simplest version, the Vasicek model prices a bond maturing at $T$ as the following average:
\be
B(t,T) = \langle \ \exp -\int_{t}^{T} du \ r(u) \ \rangle \label{bond}
\ee
where the averaging is over the possible histories of the spot rate between now
and the maturity, where the uncertainty is modelled by the noise $W$. The computation of the 
above average is
straightforward and leads to (using Eq.\ (\ref{bonddef}):
\be
f(t,\theta) = r(t) + (r_0-r(t)) (1-e^{ -\Omega \theta}) - \frac {\sigma^2}{2 
\Omega^2} (1- e^{-\Omega \theta})^2
\label{eq:vasicek_frc}
\ee

The basic results of this model are as follows:

$\bullet$ Since $\langle r_0 - r(t) \rangle =0$, the average of $f(t,\theta)-r(t)$ is given by 
\be
\langle f(t,\theta) - r(t) \rangle = -\sigma^2/2\Omega^2 (1- e^{-\Omega \theta})^2,
\ee 
and should thus be {\it negative}, at variance with empirical data. Note that in the limit $\Omega \theta \ll 1$, the order of magnitude of this (negative) term is very small: taking $\sigma=1 \%$/$\sqrt{\rm year}$ and $\theta=1$ year, it is found to be equal to $0.005 \%$ ! 

$\bullet$ The volatility $\sigma(\theta)$ is monotonously decreasing as $\exp-\Omega \theta$, while the kurtosis $\kappa(\theta)$ is identically zero (because $W$ is gaussian)

$\bullet$ The correlation function ${\cal C}(\theta)$ is proportionnal to the volatility $\sigma(\theta)$, and thus does not exhibit the marked maximum shown in Fig. 7.

$\bullet$ The variation of the spread $s(t)$ and of the spot rate should be
perfectly correlated, which is not the case (see Fig. 7): more than one factor is in any
case needed to account for the deformation of the {\sc frc}.

\begin{figure}
\centerline{\hbox{\epsfig{figure=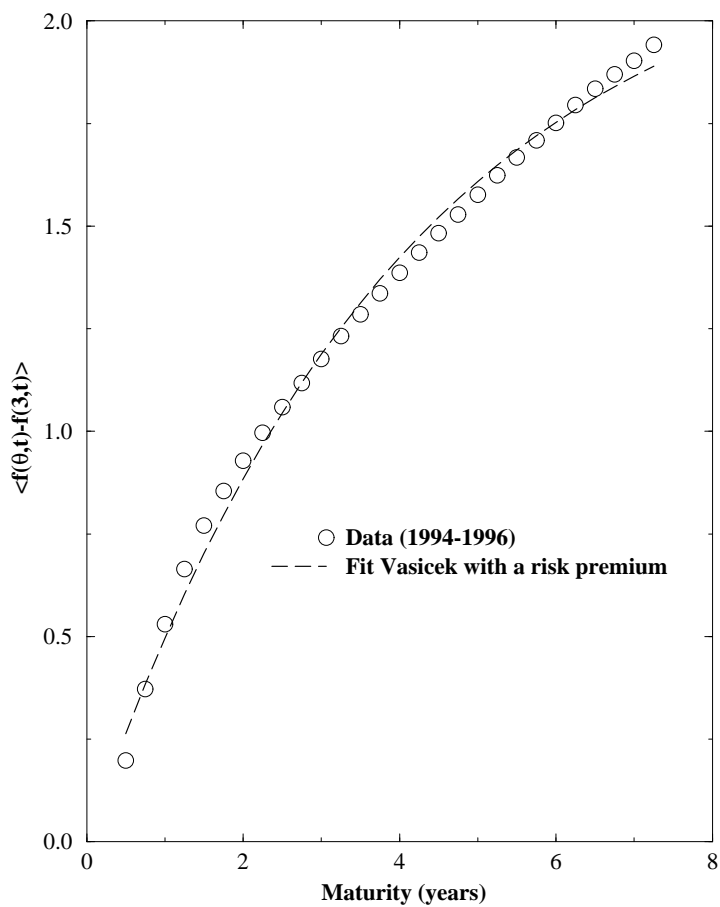,width=8cm}}}
\vskip 0.8cm
\caption{
Fit of the {\sc frc} with a Vasicek term structure, Eq.\ (\protect\ref{eq:vasicek_risk}) with a non zero market price of risk.
This is a two parameter fit -- compare to the one parameter fit in Fig. 2 -- which is however found to be acceptable. One finds $\Omega=6.17\ 10^{-2}$/year and $\lambda \sigma = 2.26\%$. Note that we neglect the 
difference between $r(t)$ and $f(t,\theta_{\min})$. }
\label{fig8}
\end{figure}

While it is easy to make $\kappa$ non zero by taking a discrete time process where
$dW$ is non gaussian, it is awkward to account for a maximum in the volatility
$\sigma(\theta)$ within such an approach (see the discussion below). However, the most obvious inconsistency of
this model is the fact that the average spread should be negative. A way out is 
to introduce the `market price of risk': as shown by Vasicek \cite{vasicek}, the probability
measure over which the average (\ref{bond}) is performed is not necessarily the historical average. Arbitrage arguments allow a `change of measure', which in the present case simply amounts to correcting the `true' (i.e historical) value of $r_0$ to an effective value $r_0 + \lambda \sigma$, where $\lambda$ is the market price
of risk. With this correction, one finds that:
\be
\langle f(t,\theta) - r(t) \rangle = \lambda \sigma (1-e^{ -\Omega \theta}) \label{eq:vasicek_risk}
\ee
A fit of the empirical data with such a formula leads to 
$\Omega=6.17\ 10^{-2}$/year and $\lambda \sigma = 2.26\%$. The value of
$\Omega$ corresponds to a mean reverting time of around $16$ years, which
is worrying since the data set is only 7 years long -- it is thus not really 
consistent to set $\langle r_0 - r(t) \rangle$ to zero. The market price of risk
$q$ corresponds to demanding for a one year maturity bond an extra return (over the spot rate) of around $0.15 \%$.

\subsection{Hull and White}

An interesting extension of Vasicek's model designed to fit exactly the `initial' 
{\sc frc} $f(t=0,\theta)$ was proposed by Hull and White \cite{hullwhite}. It amounts to replacing the
above constants $\Omega$ and $r_0$ by time dependent fonctions. For example, $r_0(t)$
represents the anticipated evolution of the `reference' short term rate itself with
time. These functions can be ajusted to fit $f(t=0,\theta)$ exactly. Interestingly,
one can then derive the following relation (for a zero market price of risk
$\lambda$):
\be
\langle \frac{\partial r(t)}{\partial t} \rangle =
\langle \left.\frac{\partial f}{\partial \theta}(t,0)\right.\rangle
\ee
up to a term of order $\sigma^2$ which turns out to be negligible, exactly for the 
same reason
as explained above. On average,
the second term (estimated by taking a finite difference estimate of the partial
derivative using the first two points of the {\sc frc}) is definitely found to be 
{\it positive}, and equal to
0.8 \%/year. On the same period (90-96), however, the spot rate has {\it decreased} 
from 8.1 \% to 5.9 \%, instead of growing by $5.6 \%$.

In simple terms, both the Vasicek and the Hull-White model mean the following: 
the {\sc frc} should basically reflect the market's expectation of the average
evolution of the spot rate (up to a correction on the order of $\sigma^2$,
but which turns out to be very small -- see above). However, since the {\sc frc} is on
average increasing with the maturity (situations when the {\sc frc} is `inverted' are
comparatively much rarer), this would mean that the market systematically expects the
spot rate to rise, which it does not. It is hard to  believe that the market persists
in error for such a long time. Hence, the upward slope of the {\sc frc} is not only
related to what the market expects on average, but that a systematic risk premium is
needed to account for this increase. Within a classical framework, this is attributed
to the `market price of risk', which is however not a directly measurable quantity.
In section (\ref{risksection}), we will propose a more direct interpretation of this
risk premium.

Let us however mention that some empirical results can be accomodated by the
Hull-White model and its extensions \cite{hullwhite2,rodgers,moraleda1}. For example, 
a `humped volatility' can be obtained by choosing $\Omega$ to be time dependent in
such a way that it starts being negative at time $t=0$ (meaning that somehow the spot
rate should {\it escape} from its reference value $r_0$) until a certain time $T_0$
beyond which it remains positive. It is easy to show in that case that the volatility
is maximum for a maturity $\theta=T_0$. However, apart from the fact that the 
interpretation of this curious shape for $\Omega(t)$ is not clear, one can also show
that in that case the average {\sc frc} should display an inflexion point at
$\theta=T_0$, which is not seen empirically. In order to fit independently the shape of the {\sc frc} and the shape
of the volatility, one should turn to  two-factor Hull-White models \cite{hullwhite2},
in which each one of the two mean reversal process can be more easily interpreted (see our own discussion below).

\subsection{Heath-Jarrow-Morton}

A second, more recent, line of thought, consists in writing stochastic differential 
equations for each of the forward rates. In this case, today's {\sc frc} is by 
construction described exactly, and provides the initial condition for the stochastic
evolution. The simplest, time translation invariant, one factor model reads
\cite{hjm,Risk}: \be
df(t,\theta)= \left[\frac{\partial f(t,\theta)}{\partial
\theta} + \mu(\theta)\right] dt + \sigma(\theta)dW(t) \label{eqhjm}
\ee
with a certain maturity dependent volatility $\sigma(\theta)$.
If the expiration date $T=t+\theta$ rather than the maturity is kept fixed, the
corresponding rate $\tilde f (t,T) \equiv f(t,\theta=T-t)$ is such that:
\be
d \tilde f(t,T) = df(t,\theta) - \frac{\partial f(t,\theta)}{\partial \theta} dt =
\mu(\theta) dt + \sigma(\theta) dW(t)
\ee
Absence of arbitrage opportunities then impose, within this continuous time
framework, a relation between $\mu$ and $\sigma$ which reads:
\be
\mu(\theta) = \sigma(\theta) \int_0^\theta \sigma(\theta') d\theta' + \lambda
\sigma(\theta), \ee
where $\lambda$ is again the market price of risk. This model has primarily
been devised to price consistently (within a no-arbitrage framework) interest rates
derivatives rather than to represent faithfully the historical evolution of the 
{\sc frc} itself. It is however interesting to remark that the above criticism still
holds if one wants to interpret data using Eq.\ (\ref{eqhjm}): since the
average value of the {\it slope} of the {\sc frc} can only be generated by the
maturity dependent drift $\mu(\theta)$, a non zero value of $\lambda$ is needed in order to
reproduce its correct order of magnitude -- the order $\sigma^2$ term is again much
too small, and can for all practical purposes be set to zero (this was first 
noticed in \cite{Campbell}). Furthermore, Eq.\ (\ref{eqhjm}) does not ensure that the shape of the {\sc frc} remains `realistic' with time: one the contrary, the {\sc frc} tends to distort more and more as time passes when the volatility has a non trivial maturity dependence.
We shall discuss in section 5 below how the introduction of a `line tension' 
may heal this disease.

In conclusion, the classical models are only consistent with empirical data provided a rather large market price of risk is included. In technical terms, this means that, in the context of arbitrage theories,  the `risk-neutral' probability measure to be used for -- say -- derivative pricing cannot be identified with the empirical probability. Other theoretical difficulties 
inherent to arbitrage theories of the interest rate curve have been discussed 
in \cite{nek2}.

\section{Risk-premium and the $\protect\sqrt{\theta}$ law}
\label{risksection}

\subsection{The average {\sc frc} and Value-at-Risk pricing}

The observation that on average the {\sc frc} follows a simple $\sqrt{\theta}$ law
(i.e. $\langle f(t,\theta) - r(t) \rangle \propto \sqrt{\theta}$) suggests a
more intuitive, direct interpretation. At any time $t$, the market anticipates either
a  future rise, or a decrease of the spot rate. However, the  average anticipated
trend is, on the long run, zero, since the spot rate has bounded fluctuations. Hence, the {\it average} market's expectation is that the future
spot rate $r(t)$ will be close to its present value $r(t=0)$. In this sense, the average {\sc frc} should thus be flat (again, up to a small $\sigma^2$
correction). However, even in the absence of any trend in the spot rate, its {\it
probable} change between now and $t=\theta$ is (assuming the simplest random
walk behaviour) of the order of $\sigma \sqrt{\theta}$, where $\sigma$ is the  
volatility of the spot rate. Hence, money lenders are tempted to protect themselves
against this potential rise by adding to their estimate of the average future rate a
{\it risk premium} of the order of $\sigma \sqrt{\theta}$ to set the forward rate at a
satisfactory value. In other words, money lenders take a bet on the future value of
the spot rate and want to be sure not to lose their bet more frequently than -- say --
once out of five. Thus their price for the forward rate is such that the probability
that the spot rate at time $t+\theta$, $r(t+\theta)$ actually exceeds $f(t,\theta)$ is
equal to a certain number $p$: 
\be \int_{f(t,\theta)}^\infty dr' P(r',t+\theta|r,t) =p
\label{VaR} \ee where $P(r',t'|r,t)$ is the probability that the spot rate is equal to
$r'$ at time $t'$ knowing that it is $r$ now (at time $t$). Assuming that $r'$
follows a simple random walk centered around $r(t)$ then leads to 
\footnote{This assumption is certainly inadequate for small times, where large
kurtosis effects are present. However, one the scale of months, these non
gaussian effects can be considered as small \cite{Livre}.}:
\be
f(t,\theta)=r(t) + A \sigma \sqrt{\theta} \qquad A=\sqrt{2}\  {\rm erfc}^{-1}(2p)\label{fA}
\ee
which indeed matches the empirical data, with $p \sim 0.16$.

At this point, we thus depart both from the so-called `unbiased expectation hypothesis' \cite{Campbell} or from Vasicek type of models, where it is assumed that
today's {\sc frc} tells us something about the {\it average} future evolution
of the spot rate, possibly corrected by an (unmeasurable) market price of risk. This,
for example, is used to calibrate the time scale $\Omega^{-1}$ which appears in the
Vasicek model, Eq.\ (\ref{eq:vasicek_frc}). In our mind, the shape of today's {\sc frc}
must rather be thought of as an envelope for the probable future evolutions of the spot
rate. The market seems to price future rates through a {\it Value at Risk} procedure
(Eqs. \ref{VaR}, \ref{fA}) rather than through an averaging procedure. In a strict
sense, Eq.\ (\ref{fA}) is not acceptable from the point of view of arbitrage, since
buying the spot rate and selling the forward rate would lead to a profit. However, one
should keep in mind that this profit is not certain because Eq.\ (\ref{fA}) only
describes the {\sc frc} in an average sense; one should also take into account the
random `deformation' of the curve described by $\xi(t,\theta)$ (which we interpret
below) which makes the above strategy
risky. Nevertheless, the simple strategy of lending long-term money and borrowing
short-term money is obviously a major source of income for most banks !

\subsection{The instantaneous {\sc frc}. The anticipated bias.}  

Let us now discuss, along the same lines, the shape of the {\sc frc} at a
{\it given instant of time}, which of course deviates from the average square root law.
As discussed in section 2,  the deviation $\xi(t,\theta)$ is actually
well approximated by the first `butterfly' mode $\xi_1(t) \Psi_1(\theta)$, which has 
the shape drawn in Fig. 4. We interpret this as follows: for a given instant of time
$t$, the market actually expects the spot rate to perform a {\it biased} random walk.
We shall argue that the market estimates the trend $m(t)$ by extrapolating the
past behaviour of the spot rate itself. However, it appears that the market
also `knows' that trends on interest rates do not persist for ever in time; the
`anticipated' bias is thus in general maturity dependent. Hence, the probability
distribution $P(r',t+\theta|r,t)$ used by the market is not centered around
$r(t)$ but rather around: 
\be r(t) + \int_t^{t+\theta}du m(t,t+u)  \ee 
where $m(t,t')$ can be called the anticipated bias at time $t'$, seen from time $t$. 

As discussed above, it is reasonable to think that the market estimates $m$ by 
extrapolating the recent past to the nearby future. Mathematically, this reads:
\be
m(t,t+u) = m_1(t) {\cal G}(u) \qquad  
m_1(t) := \int_0^\infty dv \ K(v) \delta r(t-v)
\ee
where $K(v)$ is an averaging kernel of the past variations of the spot rate.
We will call ${\cal G}(u)$ the trend persistence function; it is normalized
such that ${\cal G}(u=0)=1$, and describes how the present trend is expected to persist in th efuture. Eq.\ (\ref{VaR}) then gives: \be
f(t,\theta)=r(t) + A \sigma \sqrt{\theta} + m_1(t) \int_0^\theta du \ {\cal G}(u)
\label{3fit}
\ee
Identifying this expression with Eq.\ (\ref{mode1}), one finds that:
 
$\bullet$ As discussed above, the spread $s$ is associated to the volatility of the spot rate. Note however that other sources of risk, such as 
the exchange rate fluctuations for foreign investors, might also  be included in the `effective' volatility of the spot rate used in 
Eq.\ (\ref{VaR}).
 
$\bullet$ The persistence function is simply the derivative of the first mode $\Psi_1(\theta)$, or $\Delta(\theta)$. This function is 
plotted in Fig. 9; 
the interesting point -- related to the existence of a maximum in $\Delta(\theta)$ -- is that it becomes negative (but small) 
beyond one year. This means that the
market anticipates a trend reversion on the scale of one year.

\begin{figure}
\centerline{\hbox{\epsfig{figure=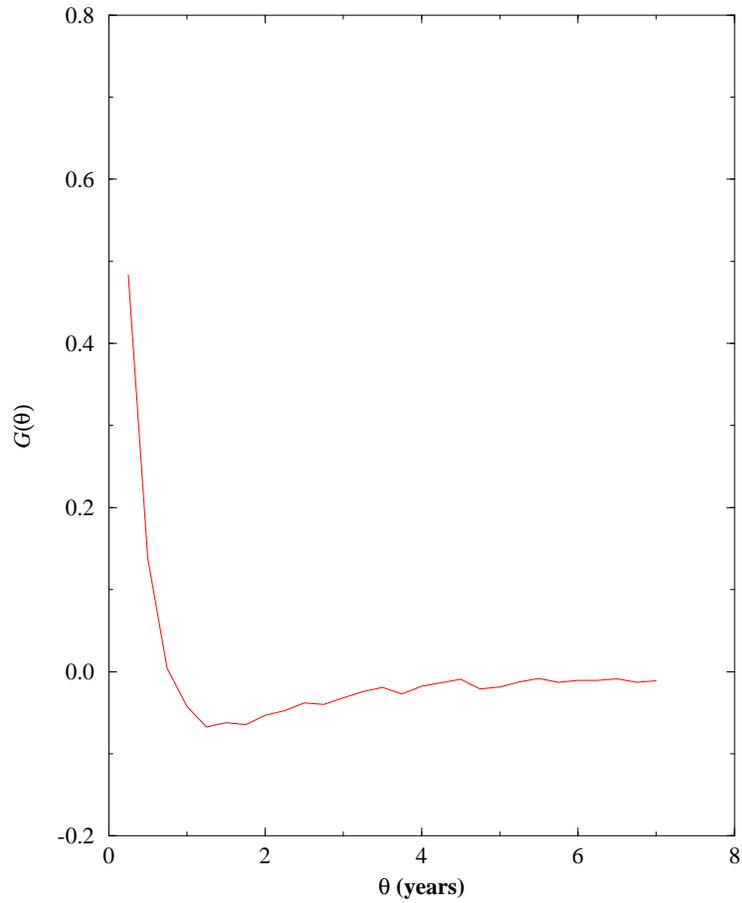,width=8cm}}}
\vskip 0.8cm
\caption{
The `persistence' function ${\cal G}(\theta)$ describing how the market 
anticipates the persistence in the future of the observed past trend of the spot rate. Note that ${\cal G}(\theta)$ becomes negative
beyond one year, meaning that the
market anticipates a trend reversion on the scale of one year.
}
\label{fig9}
\end{figure}

This interpretation furthermore allows one to understand why the three functions
introduced above, namely $\Delta(\theta)$, $\sigma(\theta)$ and the correlation function ${\cal C}(\theta)$ have similar
 shapes. Indeed, taking for simplicity the kernel $K(v)$ to be 
$\gamma \exp[-\gamma v]$, one finds:
\be
d m_1(t) = -\gamma m_1 dt + \gamma d r(t) \label{dm1}
\ee
Hence, the correlation function ${\cal C}(\theta)$ is nothing but:
\be 
{\cal C}(\theta) = \gamma \int_0^\theta du \ {\cal G}(u)
\ee
(we have assumed in the above expression that ${\cal C}(\theta_{\max}) \simeq 0$, which is not
a bad approximation -- see Fig. 7). On the
other hand, since from Eq.\ (\ref{dm1}) $\langle m_1^2 \rangle = \gamma
\sigma^2(0)/2$  one also finds:  
\be
\Delta(\theta)=\sqrt{\frac{\gamma}{{2}}} \sigma(0) \int_0^\theta du {\cal G}(u) = \frac{\sigma(0)}{\sqrt{2\gamma}} {\cal C}(\theta) 
\ee
thus showing that $\Delta(\theta)$ and ${\cal C}(\theta)$ are proportionnal. Actually, even the numerical prefactor predicted by this simplified description is quite good: using the value of  $\gamma \equiv c$ determined by the fit shown in Fig. 7, the predicted value of $\Delta(\theta^*)$ is found to be $\simeq 0.29 \%$, instead of the observed $0.38 \%$ (see Fig. 4).

Turning now to the volatility $\sigma(\theta)$, one finds that it is given by:
\be
\sigma^2(\theta)=\left[1+{\cal C}(\theta)\right]^2 \sigma^2(0) + 
\sigma_s^2 {\cal Y}(\theta)^2 \label{voltheta}
\ee
where $\sigma_s^2$ is the contribution of the spread volatility. We thus see that the
maximum of $\sigma(\theta)$ is indeed related to that of ${\cal C}(\theta)$.
Intuitively, the reason for the volatility maximum is as follows: a variation in the
spot rate changes that  market anticipation for the trend $m_1(t)$. But this change of
trend obviously has a larger effect when multiplied by a longer maturity. For
maturities beyond one year, however, the decay of the persistence function comes into
play and the volatility decreases again.

Interestingly, the form (\ref{voltheta}) suggests
that the volatility should increase again for very large $\theta$, due to the contribution of the
last term. This can indeed be seen on a restricted data set which allows to reach $\theta=10$ years (Fig. 10), where one sees  that $\sigma(\theta)$ 
reaches a {\it minimum} around $\theta=8$ years. This effect can also be related to the slow decay of the kurtosis
with $\theta$ shown in Fig. 6, which reflects the fact that short maturities are essentially 
sensitive to one factor (the spot rate fluctuations), while longer maturities are
progressively affected by a second factor (the anticipated risk) -- adding independent contributions indeed reduce the kurtosis.

\begin{figure}
\centerline{\hbox{\epsfig{figure=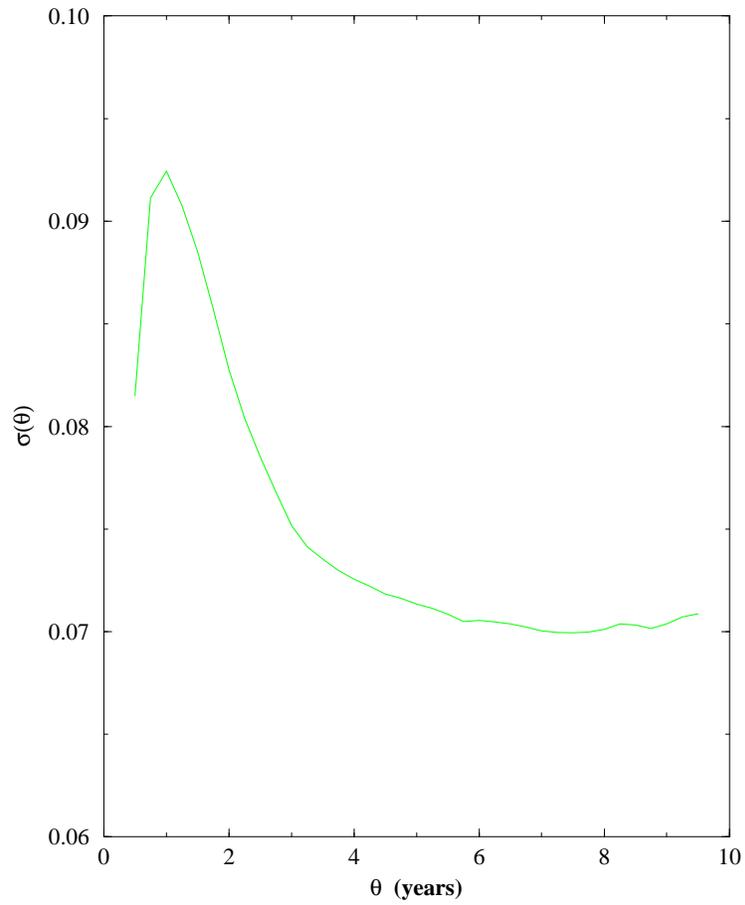,width=8cm}}}
\vskip 0.8cm
\caption{
Volatility $\sigma$ as a function of $\theta$, as in Fig. 6, but for a
restricted time period allowing to extend the range of maturities to $10$ years.
A volatility minimum appears around $\theta=8$ years, beyond which the
volatility of the spread becomes important.
}
\label{fig10}
\end{figure}

Taking for simplicity $\int_0^\theta du \ {\cal G}(u) = \theta \exp -\Gamma \theta$, as 
suggested by Fig. 7, one can now fit the instantaneous {\sc frc} by the expression
(\ref{3fit}), for example fixing $\Gamma$. One then extracts from the {\sc frc} the
anticipated bias $m_1(t)$ as a function of $t$. We find that $m_1(t)$ is correlated with the past
evolution of the spot rate, on the scale of one year (i.e. the past time scale
determining $m_1$, $\gamma^{-1}$, is comparable to the time scale of the persistence
function, $\Gamma^{-1}$).  This is perfectly compatible with the value of $\gamma=c$
extracted directly from the fit shown in Fig. 7, which gives $\gamma^{-1}=8$ months.

Finally, we want to note that Eq.\ (\ref{dm1}) is actually very similar in spirit to the
second equation in Hull and White's two factor model \cite{hullwhite2}, where the `noise' term in the
second equation is actually the spot rate itself. This model was constructed in an ad-hoc way to
reproduce the volatility hump and its intuitive meaning was not explicited. We believe that the above idea of an
`autoregressive' anticipated biais is a rather natural financial interpretation.

\section{The forward rate curve as a vibrating string}

In the previous section, we have thus argued that the shape of the instantaneous {\sc frc} is fixed by two distinct type of anticipations:

-- An anticipated bias, whose influence is expected to decay with time, and whose amplitude is determined by the recent evolution of the spot rate.

-- A anticipated risk, which gauges the potential motion of the spot rate, and which is added as a risk premium by money lenders.

Hence, as noticed by several people, the whole {\sc frc} evolves according to 
rather few `factors' \cite{NEK,Douady}, while in principle each maturity $\theta$ could
feel the influence of some independent factor. However, one also expects that
the `forces' determining the evolution of each points of the {\sc frc} act
as to prevent the {\sc frc} from `blowing apart' with time, and to diffuse the
information from one maturity to the next. A natural mechanism is the following: the
dynamical evolution should be such that large deviations  between nearby maturities are
improbable -- it is plausible that market forces tend to keep the forward rate $\tilde
f(t,T)$ close to its local average $[\tilde f(t,T-\epsilon)+f(t,T+\epsilon)]/2$, where
here $\epsilon =$ 3 months.  The simplest choice is then to write: 
\be
\delta \tilde f(t,T) = {\cal D}
\left[\tilde f(t,T-\epsilon)+\tilde f(t,T+\epsilon)-2 \tilde f(t,T) \right] + \eta(t,T)
\label{string}
\ee
where ${\cal D}$ measures the strength of this force (`line tension') which
could in principle be $\theta$ dependent. Eq.\ (\ref{string}) describes a mean
reversion of the forward rate towards its local average. It is the dynamical equation
of a string of beads connected by springs, subject to a random force term $\eta$ which
can reflect, in particular, the variations of the forward rate  which are specific to
that given maturity (for example, the influence of a major political event which is
scheduled around that particular date.). 

The term proportional to ${\cal D}$ would become, in the continuum limit, a
second order derivative $\partial^2 \tilde f/\partial T^2$, which is in principle
not allowed from arbitrage arguments, at least when the `noise' $\eta(t,T)$ contains 
a finite number of independent components. Intuitively, this can be understood in the 
simple case where $\eta(t,T)$ is independent of $T$. Then, a possible riskless winning
strategy is to buy (resp. sell) one contract of maturity $T$ if  \be \tilde
f(t,T-\epsilon)+\tilde f(t,T+\epsilon)-2 \tilde f(t,T) > 0 \ee
(resp. $<0$), and then to hedge the position by buying (or selling) the
appropriate number of the shortest maturity contract. Then, because of the
restoring `string' force, the resulting gain will be positive. The 
result of such a strategy (used simultaneously on all maturities) is shown in Fig. 11,
in the absence of transaction costs. This curve shows indeed a clear positive
slope, thereby directly showing the existence of a ${\cal D}$ term affecting the
temporal evolution of the  {\sc frc}. The fact that the above strategy is not riskless
is due to the presence of more than one random factor $\eta$. More importantly, we
have found empirically that the above growing {\sc p\&l} is completely destroyed by
transaction costs. 

Therefore, the
presence or absence of shape dependent terms (such as the above  second order
difference term) {\it cannot be decided by arbitrage arguments in the presence of transaction costs}.

\begin{figure}
\centerline{\hbox{\epsfig{figure=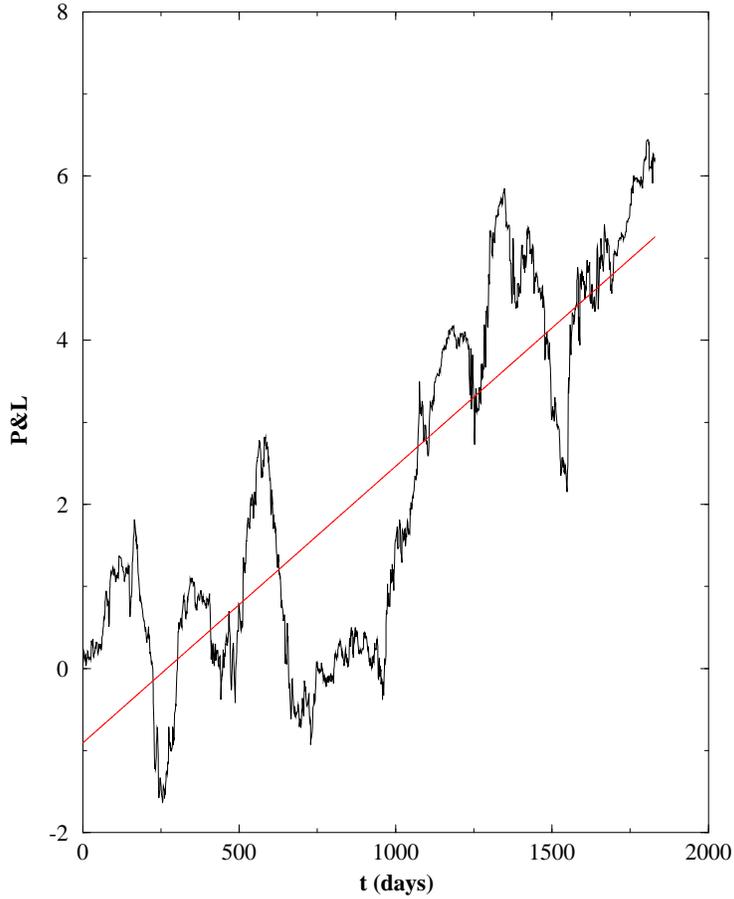,width=8cm}}}
\vskip 0.8cm
\caption{
Empirical Profit and Loss curve corresponding to the simple `arbitrage' strategy proposed in the text, which exploits the presence of a `string' force proportionnal
to the local second (discrete) derivative of the {\sc frc}. On average, this strategy clearly leads to a positive profit, although reasonable transaction costs ruins it completely. However, the presence of a non zero $\cal D$ is very important since it allows one to understand why the effective number of factors
influencing the {\sc frc} is small (see Eq. (\protect\ref{Dq2}) below).}
\label{fig11}
\end{figure}

Still, the very existence of these terms is extremely important to
understand the qualitative dynamical evolution of the {\sc frc}. In particular, the
`line tension' ensures that {\sc frc} becomes (and remains) smooth even if the
initial condition is not, of if the term structure of the volatility is not. This is not the case within Heath-Jarrow-Morton like formulations, where ${\cal D}=0$, and where temporal evolutions tend to be ill-behaved \cite{Musiela}. 

We want to show here that the mere existence of a line tension allows one to understand
qualitatively the fast decay of the eigenvalues of the correlation matrix, as observed in Fig. 5. If one introduces  $\phi(t,T)=\tilde f(t,T)-\tilde
f(t,t)$, and takes for notational simplicity the continuous time limit, one finds that
$\phi$ evolves according to: 
\be
d \phi(t,T)= {\cal D} \frac{\partial^2 \phi(t,T)}{\partial T^2} dt + 
d\eta(t,T) \qquad \phi(t,t) \equiv 0
\ee
Defining the eigenmodes of the operator $\frac{\partial^2}{\partial T^2}$ which vanish at $T=t$, i.e. : 
$\phi_q(t,T)= M_q(t) \sin (q(T-t))$, the evolution of $M_q(t)$ is then given by:
\be
\frac{\partial M_q}{\partial t}= -{\cal D} q^2 M_q + \eta_q(t)
\ee
where $\eta_q$ is the Fourier transform of the noise term. In this form
one sees that:

$\bullet$ The `wavevector' $q$ is the index labelling the eigenmodes of the matrix 
$\cal M$ introduced above.

$\bullet$ The `lifetime' of a perturbation of wavevector $q$ is
inversely
proportionnal to ${\cal D} q^2$. High $q$'s (corresponding to short
maturity differences) relax faster -- this corresponds to the fact that
if the {\sc frc} is distorted on very small scales, this will probably 
not persist in time. 

$\bullet$ The average (over time) amplitude in mode $q$, proportional to
the $q^{\rm th}$ eigenvalue of $\cal M$ is given by:
\be
\langle |M_q|^2 \rangle = \frac{\langle |\eta_q|^2 \rangle}{{\cal D}q^2}\label{Dq2}
\ee

If $\eta(t,T)$ was independent for each maturity, $\langle
|\eta_q|^2 \rangle$ would be independent of $q$, already leading to a $q^{-2}$
decay of $\langle |M_q|^2 \rangle$, due to the restoring force $\cal D$. The presence of correlations in the noise along the maturity
axis acts to make this decay even faster. For example, if the correlations 
are exponentially decaying for different $T$'s (i.e. as
$\exp-K|T-T'|$), one would find:
\be
\langle |\eta_q|^2 \rangle \propto \frac{1}{K^2+q^2}
\ee
hence leading to a much faster still decay of $\langle |M_q|^2 \rangle$. As shown in
Fig. 5, the observed decay is indeed intermediate between $q^{-2}$ and $q^{-4}$.

The main conclusion of this section is that a `line tension' term indeed 
exists in the dynamical evolution of the {\sc frc}, acting as to reduce 
the distortions along the line. This line tension is responsible for
reducing  the effective number of factors needed to interpret the shape of 
the whole {\sc frc}, even when independent shocks affect each different maturities \footnote{The idea of projecting the {\sc frc} on a few smooth
functions of the maturity can also be found in \cite{Douady,Chu}, although the basic mechanism underlying such a reduction, in terms of a line tension, was not discussed.}.
Arbitrage arguments in the absence of transaction costs  impose that ${\cal D} \equiv 0$, a conclusion that transaction costs allows one to by-pass.

\section{Summary-Conclusion}

The main results contained in this paper, based on the study of the whole
U.S. forward rate curve since 1990 (and which we confirmed on the corresponding swap
rates), are the following:

$\bullet$ The average {\sc frc} (measured from the spot rate) grows as the square-root
of the maturity, with a prefactor which is comparable to the spot rate volatility.
This strongly suggests that forward rate market prices include a risk premium,
comparable to the probable changes of the spot rate between now and maturity. This
interpretation seems to us more natural (although in the same spirit) than the one
based on an unobservable `market price of risk'.

$\bullet$ The instantaneous {\sc frc} departs form a simple square-root law. The distortion is maximum around one year, and reflects the market anticipation of the trend on the spot rate (as in more traditionnal models of interest rates \cite{Campbell}). Somewhat surprisingly, however, this trend appears to be calibrated on the past behaviour of
the spot itself: if
the spot rate has decayed over the past months, the one year forward rate is below 
the
average square root extrapolation, and vice-versa. This is consistent with the fact 
that the volatility is maximum for one year maturities.

$\bullet$ The number of independent components needed to interpret most of the 
{\sc frc} 
fluctuations is rather small, although in principle a random contribution should be 
assigned to each maturity. We rationalize this finding by showing that the dynamical
evolution of  the {\sc frc} contains a stabilizing `line tension' term, which tends to
suppress short scale distortions of the {\sc frc}. This shape dependent term in the dynamical
evolution is not usually considered in interest rate models, because it leads, in
principle, to arbitrage. However, this arbitrage cannot be implemented in practice
because of residual risks and transaction costs. This is  important because the long time statistical
properties of a line subject to random forcing is very different when the line tension
term is present (even very small) or strictly zero: the `line tension' term 
(as all second derivative terms) lead to a smoothing of the singularities, which
otherwise survive or even develop in the course of the evolution. We thus conclude that the presence
of transaction costs (or other market `imperfections', such as residual risk) is
crucial for model building, for a much wider class of models becomes eligible to
represent reality. 

The present work should be extended in several directions. First, other interest rates
should be studied, to know whether the broad features found on the U.S. market also holds
in other cases. Second, the consequences of our modelling for interest rate derivatives
should be worked out. Finally, it would be interesting to interpret the time
dependent smile in option markets in term of a `forward volatility curve', and see how the ideas expressed here transpose to this case. We hope to deal with these issues in the near future.

\vskip 1cm

Acknowledgments: We want to thank J.P. Aguilar and L. Laloux for many useful discussions.


\begin{thebibliography}{99}


\bibitem{hull}
Hull J. (1997) {\it Options, futures and other derivative securities}, Prentice-Hall.

\bibitem{Risk} see the reprints in: {\it Vasicek and beyond}, edited by Lane Hugston, Risk Publications, 1997.

\bibitem{rodgers} L.C.G. Rodgers, (1996), ``Which model for the term structure of interest rates should one use ?'', Mathematical Finance, IMA Volume 65, 93 (Springer-Verlag).


\bibitem{hjm}
Heath D., Jarrow R. \& Morton A. (1992), ``Bond pricing and the term structure of
interest rates: a new methodology for contingent claims evaluation'', {\it  Econometrica}, {\bf 60}, 77-105.

\bibitem{HoLee} 
Ho T.S.Y. and Lee S.B. (1986), ``Term structure movements and pricing Interest Rate contigent claims", {\it J. Finance} {\bf 41}, 1011


\bibitem{cir}
Cox J.C., Ingersoll J.E. \&  Ross S.A. (1981) ``The relationship between forward prices and futures prices", {\it Journal of Financial Economics}, {\bf 9}, 321-46.

\bibitem{rendleman}
Rendleman R. \& Carabini C. (1979) ``The efficiency of Treasury-bills futures markets", {\it Journal of Finance}, {\bf 34}, 895-914.

\bibitem{chan}
Chan K.C., Karolyi G.A., Longstaff F.A. \& Sanders A.B. (1992) ``An empirical
comparison of alternative models of the short term interest rates", 
{\it  Journal of Finance}, {\bf XLVII}, 3, 1209-1227.

\bibitem{Wilmott} M. Apabhai, K. Choe, F. Khennach, P. Wilmott, `Spot-on Modelling', Risk, (November 1995).



\bibitem{hullwhite2}
Hull J. \& White A. (1994) ``Numerical procedures for inplementating term 
structure models II: two-factor models'', {\it Journal of Derivatives}, 
{\bf winter},  37-48.

\bibitem{moraleda2}
Moraleda J.M.  (1997) ``On the pricing of interest rate options'', PhD Thesis, Erasmus University, Rotterdam.

\bibitem{Longstaff} F. Longstaff, E. Schwartz (1992), ``Interest rate volatility and the term structure: a two factor equilibrium model'', Journal of Finance, {\bf 47}, 1259.



\bibitem{schlogl}
Schl\"ogl E. \& Sommer D. (1997) ``Factor models and the shape of the term structure" AFFI Conference, Grenoble.



\bibitem{Douady} R. Douady, ``Yield curve smoothing and residual variance of fixed income positions'', Working paper, June 1997.

\bibitem{vasicek}
Vasicek O.A. (1977) ``An equilibrium characterization of the term structure", {\it  Journal of Financial Economics}, {\bf 5}, 177-188.



\bibitem{hullwhite}
Hull J. \& White A. (1993), ``One factor interest rate models and the valuation
of interest rate derivative securities'', {\it Journal of Financial and Quantitative Analysis}, {\bf 28},  235-254.

\bibitem{moraleda1}
Moraleda J.M. \& Pelsser A. (1997) ``Forward vs Spot Interest-rate models of the term-structure: an empirical comparison",  Working Paper, Erasmus University, Rotterdam.

\bibitem{nek2} N. El Karoui, A. Frachot, H. Geman, ``A note on the behaviour of long zero coupon rates in no arbitrage framework'', Working Paper, May 1994.

\bibitem{Livre} J.P. Bouchaud, M. Potters, {\it Th\'eorie des Risques Financiers}, Al\'ea-Saclay/Eyrolles (1997). 

\bibitem{Campbell} J.Y. Campbell (1986), ``A defense of traditional hypotheses about the term structure of interest rates'', Journal of Finance, XLL, March.
See also J.Y. Campbell, A.W. Lo, C. McKinlay (1997), {\it Econometrics of Financial Markets}. 


\bibitem{NEK} see e.g. N. El Karoui, H. Gemam, V. Lacoste, ``On the role of State Variables 
in Interest Rate Models'', Working Paper (Universit\'e Paris VI), January 1996.

\bibitem{Musiela} M. Musiela, ``On the structure of PDEs arising in Interest Rate Models'', talk given in Paris, November 1997.

\bibitem{Chu} M. Chu, ``The Random Yield Curve and Interest Rate Options'', Working Paper (November 1996).

\end{thebibliography}
\end{document}